# Study on MHD activity in Tokamaks

Fatemeh Hajakbari[1,2] and Alireza Hojabri[1,2]

[1]*Physics group, Islamic Azad University of Karaj, Iran.*
[2]*Plasma Physics Research Center, Azad University, Tehran 14835-197,Iran*

The tearing mode instability and the associated magnetic islands can lead to a degradation of tokamak plasma performance and eventually to a disruption. A crucial issue for the extension of advanced tokamak scenarios to long pulse operation is to avoid these MHD instabilities. Prevention of disruptions and limitations of adverse effect on energy confinement depend on the avoidance of formation of magnetic islands, and it depends also on the successful control of the growth modes excited by intrinsic plasma effects such as the neo-classical drive of unfavourable bootstrap current gradient at some q rational surfaces. In this paper are described in the influence of mode coupling on the nonlinear evolution of rotating magnetic islands.

**Introduction**

The purpose of this paper is to investigate further the major disruption in low-q discharge in IR-T1 and comparison between the theoretical and experimental results rate of island growth. Experimantal evidence has shown that disruptions are connected with MHD activity and that the time-scale involoved is intermediate between the resistive and ideal MHD time-scale [1]. The explosive growth of these islands leads to their overlapping and the field lines become stochastic. These growth time values span the range between those expected from ideal and resistive MHD time-scale. The time scale for ideal mode growth is generally bounded from below by the ideal time scale and from above by the resistive time-scale $\left(\tau_A \prec \tau_g \prec \tau_S\right)$.

Typical ideal time scales are given by the poloidal Alfven time $\tau_A \equiv \frac{a}{V_A}$ where $V_A$ is the Alfven velocity and resistive time scale [2]:

$$\tau_S \cong \tau_R^{0.6} \times \tau_A^{0.4} \quad (1)$$

where $\tau_R \equiv \frac{\mu_o a^2}{\eta(o)}$ is the resistive skin time, where $\eta(o)$ is the resistive at the plasma center.

**Experimental Results and Discussion**

The IR-T1 is a conventional tokamak with a major and minor radii of 45cm and 12.5cm, respectively, and a circular cross section without a copper shell and divertor and using a material limiter of minor radius 11.5cm. The plasma parameter in this work is about:

Plasma current: $30kA < I_p < 60kA$

Toroidal field: $6kG < B_t < 9kG$

Central electron temperature: $50eV < T_e < 200eV$

Confinement time: $0.5ms \leq \tau_e \leq 3ms$

Impurity level: $1 < Zeff < 2.5$

In figure1 we shows the evolution of plasma current, loop voltage, horizontal displacement, MHD activity and ECE signal before and during the disruptive instability at time 12-17ms. The approximately explosive growth of the precursor activity is presented at time 12.6ms correlated with decrease temperature that observed in ECE signal (fig.1.f) and increase $H_\alpha$ radiation (fig.1.e).

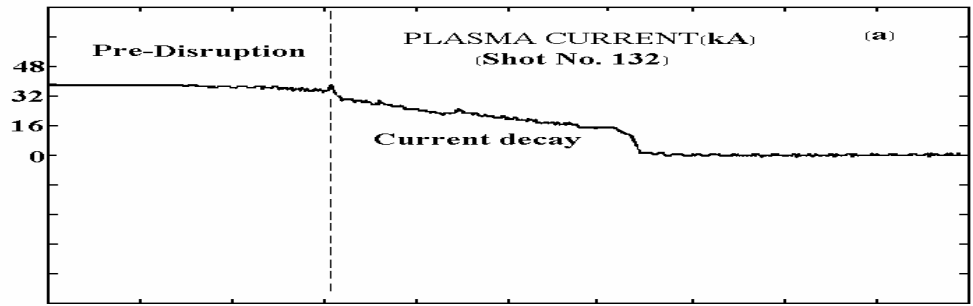

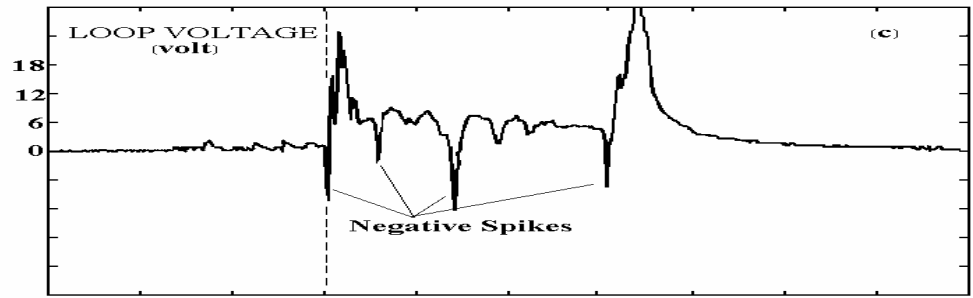

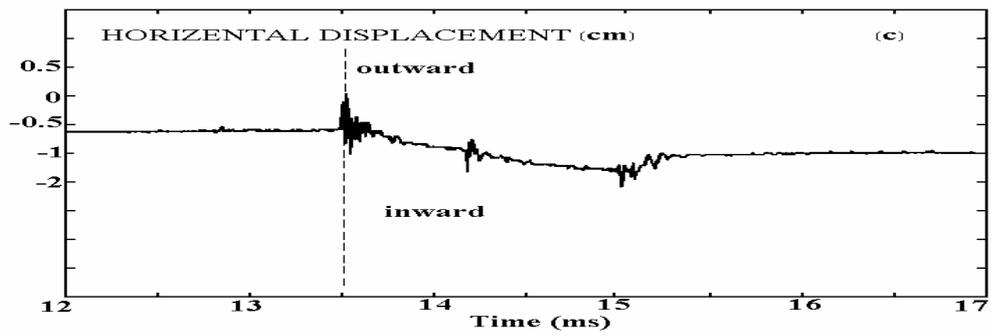

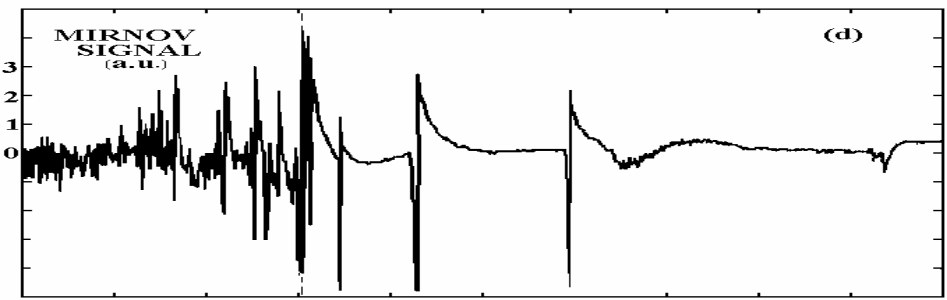

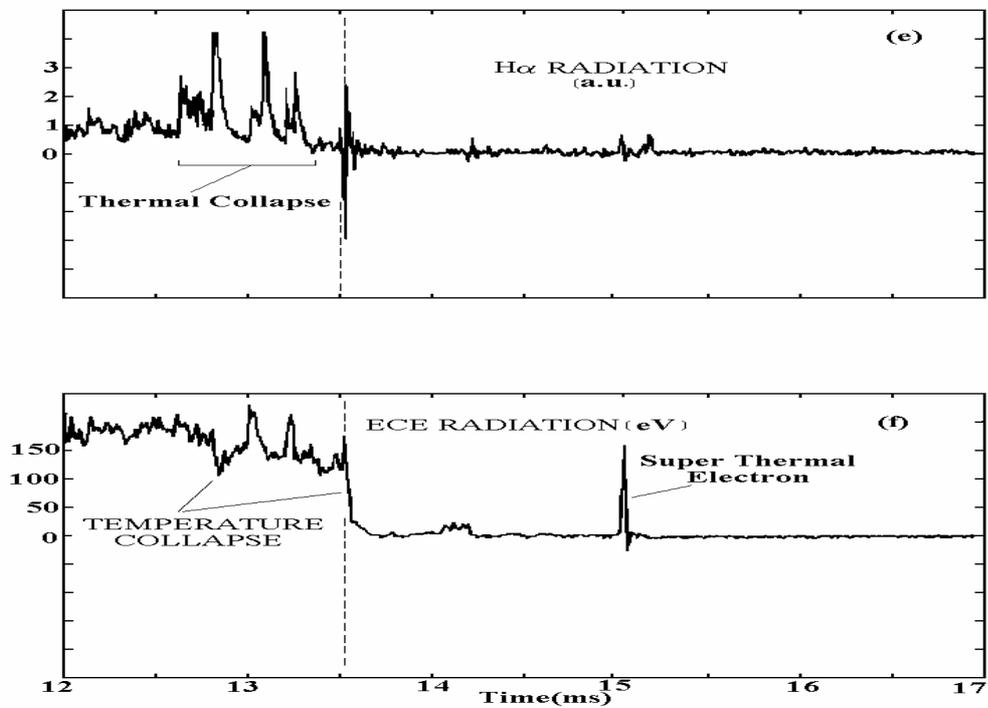

Figure 1. Plasma current (a), loop voltage (b), horizontal displacement signal (c), Mirnov coil signal (d), $H_\alpha$ signal (e), ECE signal (f) in expanded time-scale (12-17ms) before and during the disruptive instability.

**Conclusion**

The magnitude of tokamak m=2 poloidal fluctuations (Mirnov oscillations) dependence on the value of the safety factor at the limiter. We observed that as a function of q(a), the amplitude of the m=2 mode exhibits a maximum in IR-T1.

**References**

[1] JAHNS, G. L., et al., *Nucl. Fusion,* **28** (1988) 881.

[2] SCHULLER, F.C., *Plasma Phys. Control Fusion* **37**(1995) A135.

[3] Ghoranneviss M., Hojabri A. and Doranian D., *Plasma Fusion Res*. **3** (2000) 214.